\documentclass[aps,prl,showpacs,twoside,twocolumn,10pt]{revtex4-2}
\usepackage[colorlinks=true, citecolor=red, urlcolor=blue ]{hyperref}
\usepackage{times,epsfig,amssymb,amsfonts,amsmath,bm,subfigure,mathtools,amsthm,braket,soul,enumitem,xcolor,physics,graphics,graphicx}
\UseRawInputEncoding
\usepackage[normalem]{ulem}
\usepackage{comment}
\usepackage{epstopdf}

\usepackage{booktabs}

\begin{document}

\title{Property-Guided Diffusion for Inverse Design of Crystalline Materials}

\author{Sourav Mal$^{1,3}$, Subhankar Mishra$^{2,3}$ Prasenjit Sen$^{1,3}$}
\affiliation{$^1$Harish-Chandra Research Institute, Chhatnag Road, Jhunsi, Prayagraj 211019, India}
\affiliation{$^2$School of Computer Sciences, National Institute of Science Education and Research (NISER), Jatni, Odisha, 752050, India}
\affiliation{$^3$Homi Bhabha National Institute, Training School Complex, Anushakti Nagar, Mumbai, 400094, India}

\begin{abstract}
Diffusion-based generative models with property guidance have emerged as a promising paradigm for inverse materials design by enabling the generation of crystalline materials with user-specified target properties. However, despite recent advances, the effectiveness of property guidance, its influence on crystallographic symmetry, and the physical viability of generated materials remain poorly understood. To address these questions, we develop a property-guided framework based on the lightweight diffusion model DiffCrysGen using parameter-efficient adapter fine-tuning and classifier-free guidance (CFG). The resulting framework enables efficient multi-property crystal generation while preserving the knowledge learned during unconditional pre-training. Using formation energy together with saturation magnetization and Vickers hardness as representative inverse-design tasks, we systematically investigate the influence of CFG across a broad range of guidance strengths. Increasing the guidance scale progressively steers the generated property distributions toward the prescribed targets while reducing the fraction of lowest-symmetry ($P1$) structures and increasing the proportion of higher-symmetry structures. To evaluate physical viability, generated structures are geometrically prescreened and subsequently validated using a machine-learning interatomic potential (MLIP)-based workflow comprising structural relaxation and thermodynamic, dynamical, and property-specific analyses. The framework identifies thermodynamically and dynamically stable magnetic and mechanically hard materials with overall success rates of 12.3\% and 3.9\%, respectively. These results establish property-guided DiffCrysGen as an efficient framework for inverse materials design while providing new insights into the role of classifier-free guidance in crystal generation.
\end{abstract}

\maketitle

\section{Introduction}

The discovery of crystalline materials with tailored functional properties is fundamental to modern materials science, enabling advances in technologies ranging from permanent magnets and semiconductors to catalysts, batteries, superconductors, and quantum devices. Although first-principles electronic-structure calculations have become indispensable for accurately predicting properties of crystalline materials, their computational cost makes exhaustive exploration of the vast crystal chemical and structural design space prohibitively expensive~\cite{materials_project, aflow, Curtarolo2013, OQMD}. This challenge has motivated the emerging paradigm of inverse materials design~\cite{Zunger2018InverseDesign}, which seeks to identify previously unknown crystal structures that satisfy predefined target properties, rather than predicting the properties of existing materials. Realizing this vision, however, remains a formidable challenge because the relationship between crystal structure and functional properties is highly complex, while the number of accessible crystalline materials is effectively unbounded. Recent advances in generative artificial intelligence have transformed computational materials discovery by enabling the \emph{de novo} generation of crystal structures, thereby offering an attractive alternative to exhaustive exploration of crystal design space~\cite{pascal_friederich, Cheng2026, DeBreuck2025, Chen2025}. 

Among generative approaches, diffusion models~\cite{ddpm, song2021} have emerged as the state-of-the-art framework for crystal generation owing to their ability to generate diverse, crystallographically valid, and chemically realistic materials. Recent developments have broadened the scope of crystal diffusion models, including unconditional crystal generation~\cite{cdvae, diffcsp, diffcrysgen}, symmetry-aware crystal generation through explicit crystallographic constraints~\cite{diffcsp++, symmcd, wyckoffdiff}, property-conditioned inverse design~\cite{mattergen}, text-guided inverse design~\cite{chemeleon, tgdmat}, and reinforcement-learning-based fine-tuning of diffusion models~\cite{chemeleon2, matinvent}. Collectively, these advances have transformed diffusion-based crystal generation from an unconditional generative task into a versatile framework for inverse materials design, enabling the generation of materials conditioned on text-based descriptions or target properties.

Despite these recent advances, several fundamental challenges remain. Existing text-guided frameworks~\cite{chemeleon,tgdmat} enable flexible generation from natural-language descriptions but do not provide direct control over continuous target properties. On the other hand, models optimized explicitly for property-targeted generation employ either reinforcement learning (RL) alignment~\cite{chemeleon2,matinvent} or inference-time property guidance~\cite{mattergen}. While these approaches have demonstrated inverse design for specific design targets, they leave several important methodological gaps. RL-based methods permanently modify the generative distribution during training, making the conditioning behaviour fixed after optimization rather than adjustable at inference time. By contrast, guidance-based methods have devoted comparatively little attention to understanding the generative behaviour induced by the guidance mechanism itself. In particular, it remains unclear how classifier-free guidance influences target controllability, crystallographic symmetry, structural diversity, and the physical viability of the generated materials following rigorous atomistic validation. Furthermore, practical inverse materials design requires generating and screening millions of candidate materials, yet existing conditional diffusion models remain computationally expensive and exhibit relatively slow sampling throughput, limiting their applicability to high-throughput discovery campaigns.

To address these challenges, in this work we develop a property-guided diffusion framework based on DiffCrysGen~\cite{diffcrysgen}, a lightweight unconditional diffusion model originally developed for high-throughput crystal generation. DiffCrysGen was selected because its computational efficiency provides substantially higher sampling throughput than existing diffusion-based crystal generators while maintaining competitive generation quality. The proposed framework combines parameter-efficient adapter fine-tuning with classifier-free guidance (CFG)~\cite{cfg}, enabling multi-property inverse design while maintaining the robust generative capabilities of the pre-trained diffusion model. Using two representative inverse-design tasks involving formation energy together with saturation magnetization and Vickers hardness, respectively, we systematically investigate how guidance strength governs target convergence, crystallographic symmetry, structural diversity, and the physical viability of the generated materials, thereby establishing practical design principles for controllable crystal generation.

In addition, we establish a comprehensive validation workflow that integrates machine learning interatomic potential (MLIP)-based structural relaxation with thermodynamic, dynamical, and property-specific analyses to rigorously evaluate the physical viability of generated materials. By combining controllable crystal generation with rigorous atomistic validation, our framework provides an efficient route toward scalable inverse materials design. We anticipate that this approach will facilitate high-throughput discovery of functional crystalline materials across a broad range of technological applications.

\section{Methods}

\subsection{Fine-tuning the score network with property-specific adapter modules}

DiffCrysGen was originally trained as an unconditional diffusion model on a large corpus of inorganic crystalline materials, enabling it to learn a robust score function associated with the underlying data distribution. To transform this unconditional model into a property-guided inverse-design framework, the score network must additionally learn property-conditioned score functions from labeled datasets. However, datasets containing related quantum-mechanical properties are typically much smaller than those used during unconditional pre-training, making full retraining computationally expensive and prone to overfitting. To address this challenge, we adopt the adapter-based fine-tuning strategy introduced in MatterGen~\cite{mattergen} and integrate it into the U-Net~\cite{unet} score network of DiffCrysGen.

\begin{figure}[htbp]
\centering
\includegraphics[width=0.45\textwidth]{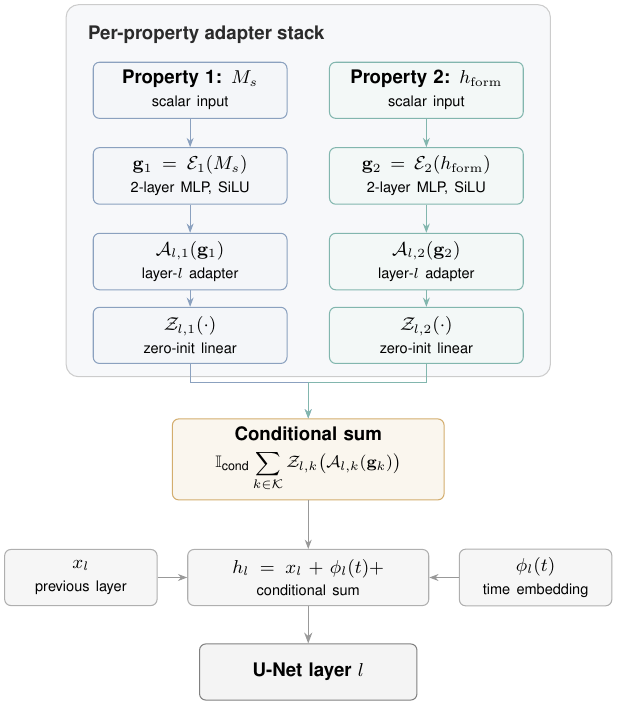}
\caption{\textbf{Architecture of the property-guided DiffCrysGen framework.}
For each conditioning property (illustrated here for saturation magnetization $M_s$ and formation energy $h_{\mathrm{form}}$), the scalar input is transformed into a latent property embedding using a two-layer multilayer perceptron (MLP) with SiLU activations. The embedding is subsequently processed by a layer-specific adapter and projected through a zero-initialized linear mix-in layer before being aggregated across all conditioning properties. The resulting conditional representation is combined additively with the intermediate feature representation $x_l$ and the diffusion timestep embedding $\phi_l(t)$ to form the input $h_l$ of the $l$-th U-Net layer. The architecture naturally accommodates arbitrary combinations of conditioning properties through independent property-specific adapter branches.}
\label{fig:arch}
\end{figure}

Specifically, lightweight trainable adapter modules are inserted into every layer of the pre-trained U-Net. Each adapter comprises a two-layer multilayer perceptron (MLP) followed by a zero-initialized linear mix-in layer. Initializing the mix-in weights to zero ensures that, at the beginning of fine-tuning, the network exactly reproduces the unconditional score function learned during pre-training. Consequently, the model retains the unconditional representations learned during pre-training while progressively adapting the score network to incorporate property-dependent information.
Fig.~\ref{fig:arch} provides a schematic overview of the proposed property-guided conditioning framework and the integration of the adapter modules into the U-Net score network.

Our implementation is modular and naturally accommodates arbitrary combinations of conditioning properties. Rather than modifying the network architecture for each new inverse-design task, property-specific adapter branches are instantiated according to the supplied conditioning properties. Let $\mathcal{P}=\{p_k\}_{k\in\mathcal{K}}$
denote the set of conditioning properties, where $\mathcal{K}$ is the corresponding index set and each $p_k\in\mathbb{R}$ is a continuous scalar property. For each property $p_k$, we first compute a property embedding $\mathbf{g}_k$ using a dedicated embedding network $\mathcal{E}_k$,

\begin{equation}
    \mathbf{g}_k = \mathcal{E}_k(p_k)
    = \mathbf{W}_{\mathcal{E}2,k}
    \left[
    \mathrm{SiLU}
    \left(
    \mathbf{W}_{\mathcal{E}1,k}p_k
    +
    \mathbf{b}_{\mathcal{E}1,k}
    \right)
    \right]
    +
    \mathbf{b}_{\mathcal{E}2,k},
\end{equation}

where $\mathbf{W}$ and $\mathbf{b}$ denote the trainable weights and biases of the embedding network, projecting the scalar property into a high-dimensional latent representation.

For every U-Net layer $l$, the property embedding $\mathbf{g}_k$ is subsequently processed by a layer-specific adapter network $\mathcal{A}_{l,k}$, allowing the conditional information to be adapted to the corresponding level of feature abstraction. Similar to the embedding network, each adapter is implemented as a two-layer MLP with a SiLU activation,

\begin{equation}
    \mathcal{A}_{l,k}(\mathbf{g}_k)
    =
    \mathbf{W}_{\mathcal{A}2,l,k}
    \left[
    \mathrm{SiLU}
    \left(
    \mathbf{W}_{\mathcal{A}1,l,k}\mathbf{g}_k
    +
    \mathbf{b}_{\mathcal{A}1,l,k}
    \right)
    \right]
    +
    \mathbf{b}_{\mathcal{A}2,l,k}.
\end{equation}

Immediately before entering the $l$-th U-Net layer, the conditional information is injected into the hidden representation by summing the contributions from all conditioning properties,

\begin{equation}
    h_l
    =
    x_l
    +
    \phi_l(t)
    +
    \mathbb{I}_{\mathrm{cond}}
    \sum_{k\in\mathcal{K}}
    \mathcal{Z}_{l,k}
    \left(
    \mathcal{A}_{l,k}(\mathbf{g}_k)
    \right),
\end{equation}

where $x_l$ denotes the intermediate crystal feature representation at the $l$-th layer of the U-Net score network, while $\phi_l(t)$ represents the layer-specific embedding of the diffusion timestep $t$, encoding the current noise level in the reverse diffusion process. The indicator function $\mathbb{I}_{\mathrm{cond}}\in\{0,1\}$ is used to implement classifier-free guidance (CFG)~\cite{cfg} by suppressing all conditional signals during unconditional training or sampling. The operator $\mathcal{Z}_{l,k}$ denotes the zero-initialized linear mix-in layer,

\begin{equation}
    \mathcal{Z}_{l,k}(\mathbf{v})
=
\mathbf{W}_{\mathcal{Z},l,k}\mathbf{v},
\end{equation}
where $\mathbf{W}_{\mathcal{Z},l,k}=\mathbf{0}$ at initialization and the projection dimension matches the channel dimension of $x_l$.

During fine-tuning, all network parameters, including both the original U-Net weights and the newly introduced adapter modules, are optimized jointly. By initializing from the pre-trained unconditional score network, the model efficiently transfers the structural knowledge learned during large-scale pre-training while adapting to property-conditioned generation using comparatively small labeled datasets. This strategy enables sample-efficient fine-tuning of DiffCrysGen across diverse inverse-design tasks without retraining the model from scratch.


\subsection{Model Training and Optimization}

The conditional model was initialized from the pre-trained unconditional model. Specifically, the weights of the original U-Net score network were transferred directly from the pre-trained model, whereas the newly introduced property-specific adapter modules were initialized randomly. This initialization strategy enables the model to leverage the structural representations learned during large-scale unconditional pre-training while efficiently adapting to property-conditioned generation.

CFG was implemented using conditioning dropout during training. For each mini-batch, the conditioning information was randomly removed with probability $p_{\mathrm{uncond}}=0.2$, allowing the model to learn both conditional and unconditional score functions within a single network. The same conditioning dropout strategy was also applied during validation to ensure consistent evaluation.

The conditional model was optimized using the same denoising score-matching loss as the original unconditional DiffCrysGen model. The only difference is that the score network $D_{\theta}$, where $\theta$ denotes the trainable parameters, now additionally takes the conditioning properties as input,
\[
D_{\theta}(\mathbf{x}_t,t)
\rightarrow
D_{\theta}(\mathbf{x}_t,t,\mathcal{P}).
\]

The model was trained for 200 epochs using the Adam optimizer with zero weight decay and an initial learning rate of $5\times10^{-4}$. A manually scheduled step-decay learning-rate policy was adopted, reducing the learning rate to $1\times10^{-4}$ after 50 epochs and subsequently to $5\times10^{-5}$ after 150 epochs. Model performance was evaluated on the validation set at the end of every epoch, and the checkpoint achieving the lowest validation loss was selected for all subsequent generation experiments.

\subsection{Property-guided crystal generation with classifier-free guidance}

Following fine-tuning, controllable crystal generation is performed using CFG during reverse diffusion sampling. Consistent with the original CFG formulation~\cite{cfg}, the same score network is used to compute both conditional and unconditional scores by simply enabling or disabling the conditioning pathway through the indicator function $\mathbb{I}_{\mathrm{cond}}$. 

At diffusion timestep $t$, let $\mathbf{x}_t$ denote the noisy crystal representation and $\mathcal{P}$ the set of conditioning properties. The conditional score is obtained by enabling the conditioning pathway ($\mathbb{I}_{\mathrm{cond}}=1$),

\begin{equation}
D_{\mathrm{cond}}
=
D_{\theta}(\mathbf{x}_{t},t,\mathcal{P}),
\end{equation}

whereas the unconditional score is obtained by disabling the conditioning pathway ($\mathbb{I}_{\mathrm{cond}}=0$),

\begin{equation}
D_{\mathrm{uncond}}
=
D_{\theta}(\mathbf{x}_{t},t,\emptyset).
\end{equation}

CFG steers the reverse diffusion trajectory towards the desired target properties by extrapolating the conditional score away from the unconditional score according to

\begin{equation}
D_{\mathrm{CFG}}
(\mathbf{x}_{t},t,\mathcal{P})
=
D_{\mathrm{uncond}}
+
w
\left(
D_{\mathrm{cond}}
-
D_{\mathrm{uncond}}
\right),
\end{equation}
where $w\ge0$ denotes the guidance scale. Setting $w=0$ recovers unconditional generation, $w=1$ corresponds to the conditional prediction, and larger values increasingly steer the reverse diffusion process towards the specified target properties, typically at the expense of structural diversity.

The guided score $D_{\mathrm{CFG}}$ is subsequently employed within the sampling framework proposed by Karras \textit{et al.}~\cite{karras2022edm}. Starting from Gaussian noise, the crystal representation is iteratively denoised over a sequence of noise levels

\begin{equation}
\sigma_i
=
\left(
\sigma_{\max}^{1/\rho}
+
\frac{i}{N-1}
\left(
\sigma_{\min}^{1/\rho}
-
\sigma_{\max}^{1/\rho}
\right)
\right)^{\rho},
\end{equation}

where $N=100$ denotes the total number of sampling steps, with $\sigma_{\max}=80$, $\sigma_{\min}=0.002$, and $\rho=7$ controlling the non-linear spacing of the noise schedule. At each sampling step, the crystal representation is updated using a second-order predictor--corrector solver consisting of an Euler predictor followed by a Heun correction. After the final denoising step, the normalized crystal representation is transformed back to the physical crystal structure using the inverse min--max transformation, yielding the generated lattice parameters, atomic species, and fractional atomic coordinates.

\begin{figure*}[htbp]
\centering
\includegraphics[width=1.04\textwidth]{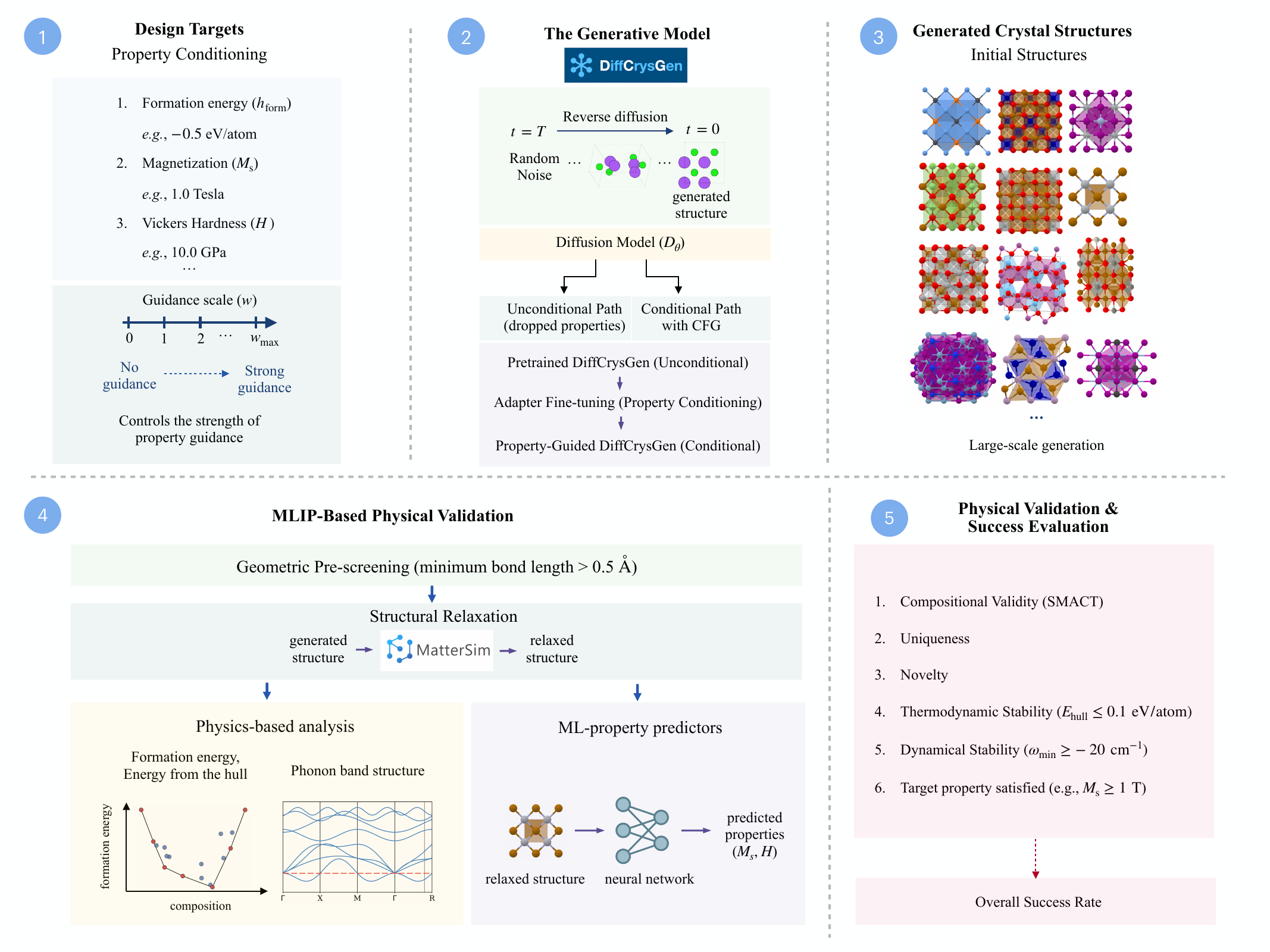}
\caption{\textbf{Schematic overview of the proposed property-guided crystal generation and evaluation workflow.} User-specified target properties, such as formation energy ($h_{\mathrm{form}}$) and saturation magnetization ($M_s$), are incorporated into the unconditional diffusion model DiffCrysGen through property-specific adapter modules, while sampling is performed using classifier-free guidance (CFG). The guidance scale $w$ controls the strength of property conditioning during the reverse diffusion process. The conditional model generates candidate crystal structures, which are first geometrically prescreened and subsequently refined using MatterSim, a machine-learning interatomic potential (MLIP). The optimized structures are then subjected to a physical validation pipeline comprising compositional, thermodynamic, dynamical, and target-property assessments, based on which different metrics have been introduced to evaluate the entire inverse-design framework.}
\label{fig:schematic-workflow}
\end{figure*}

\subsection*{Physical validation and evaluation protocol}

To assess the physical viability of the generated crystal structures, we employed a machine-learning interatomic potential (MLIP)-based validation workflow using the MatterSim universal foundation model (MatterSim-v1.0.0-5M)~\cite{mattersim}. MatterSim was interfaced through the Atomic Simulation Environment (ASE)~\cite{ase}. This workflow enables efficient structural relaxation together with the evaluation of crystallographic symmetry, thermodynamic stability, dynamical stability, and structural fidelity. 

Before structural relaxation, the generated structures were subjected to a computationally inexpensive geometric pre-screening. Structures containing any interatomic distance smaller than $0.5$~\AA\ were discarded, as such configurations are physically unrealistic, and the remaining candidates were subsequently processed through the MLIP-based validation workflow. 

The structures were fully relaxed using the FIRE optimizer together with the Fr\'echet cell filter (\texttt{FrechetCellFilter}), allowing simultaneous optimization of the atomic coordinates, lattice vectors, cell shape, and cell volume. Structural optimization was performed for a maximum of 1000 steps and was considered converged when the maximum residual force satisfied $f_{\max}\le0.005~\mathrm{eV\,\AA^{-1}}$. The crystallographic space group was determined both before and after relaxation using \texttt{spglib}~\cite{spglib} with a symmetry tolerance of $\texttt{symprec}=0.1$~\AA.

The formation energy ($h_\mathrm{form}$) of each relaxed structure was computed from the MatterSim total energy using elemental reference energies obtained with MatterSim. The corresponding energy above the convex hull ($E_\mathrm{hull}$) was then evaluated using the \texttt{PhaseDiagram} module in pymatgen~\cite{pymatgen} by constructing a phase diagram from Materials Project entries~\cite{materials_project} together with the MLIP-predicted $h_\mathrm{form}$ of the generated candidate. MatterSim predicts $h_\mathrm{form}$ with a reported mean absolute error of 25~meV/atom relative to DFT, making the resulting $E_\mathrm{hull}$ values well suited for high-throughput stability screening of generated materials.

The dynamical stability of the relaxed structures was evaluated using harmonic phonon calculations implemented in ASE~\cite{ase}. Force constants were obtained using the finite-displacement method with a $3\times3\times3$ supercell and an atomic displacement amplitude of $0.05$~\AA, where the atomic forces for each displaced configuration were evaluated using MatterSim. After constructing the force-constant matrix, the acoustic sum rule was enforced to ensure that the three acoustic phonon modes vanish at the $\Gamma$ point. Phonon band structures were subsequently evaluated along the standard high-symmetry Brillouin-zone path automatically generated by ASE from the relaxed Bravais lattice. The minimum phonon frequency ($\omega_{\mathrm{min}}$) was used as the descriptor to decide dynamical stability,  where negative values conventionally represent imaginary phonon modes and therefore indicate dynamical instability.

The structural fidelity of the generated crystals was quantified by calculating the root-mean-square displacement (RMSD) between the generated and MLIP-relaxed structures using the \texttt{StructureMatcher} implementation in pymatgen. Structure matching employed lattice tolerance $\texttt{ltol}=0.5$, site tolerance $\texttt{stol}=0.5$, angle tolerance $\texttt{angle\_tol}=10^\circ$, with supercell matching enabled (\texttt{attempt\_supercell=True}) and lattice scaling disabled (\texttt{scale=False}). Since the RMSD reported by \texttt{StructureMatcher} is normalized by $(V/N)^{1/3}$, where $V$ is the unit-cell volume and $N$ is the number of atoms, the reported RMSD values were converted back to absolute displacements in units of \AA\ for subsequent analysis.

The optimized structures obtained from this MLIP workflow were subsequently used to evaluate crystallographic symmetry, structural fidelity as measured by RMSD, thermodynamic stability, and dynamical stability. For the generated (unrelaxed) crystal structures, the target properties, namely formation energy ($h_\mathrm{form}$), saturation magnetization ($M_\mathrm{s}$), and Vickers hardness ($H$), were evaluated using surrogate machine-learning property prediction models. The $h_\mathrm{form}$ and $M_\mathrm{s}$ predictors were adopted from our previous work~\cite{diffcrysgen}, whereas the $H$ predictor was developed in the present work using the same neural network architecture. These models achieve mean absolute errors (MAEs) of $0.046$~eV/atom, $0.054$~T, and $2.549$~GPa for $h_\mathrm{form}$, $M_\mathrm{s}$, and $H$, respectively. Following MLIP structural relaxation, the thermodynamic stability of the optimized structures was evaluated directly from the MatterSim total energies, while the magnetic and mechanical properties ($M_\mathrm{s}$ and $H$) were predicted using the corresponding surrogate models. The overall workflow adopted in this work is summarized in Fig.~\ref{fig:schematic-workflow}. 

\subsection*{Evaluation metrics and success criteria}

The geometrically prescreened candidates were evaluated using progressively more stringent compositional, structural, thermodynamic, dynamical, and functional criteria. Composition validity was assessed using SMACT~\cite{smact}, ensuring charge-balanced and chemically plausible compositions. Uniqueness was defined as the absence of duplicate generated structures, whereas novelty was determined by comparing the generated structures against the MP-Alex reference database using the ordered--disordered structure-matching scheme proposed in MatterGen~\cite{mattergen}. Thermodynamic stability was defined by $E_{\mathrm{hull}}\le0.1$~eV/atom, whereas dynamical stability was defined by $\omega_{\mathrm{min}}\ge-20~\mathrm{cm}^{-1}$, allowing a small tolerance for numerical uncertainties arising from finite-displacement phonon calculations, finite supercell sizes, and the approximate nature of MLIP-predicted forces. Finally, the generated candidates were required to satisfy the prescribed target properties, namely $M_\mathrm{s}\ge1.0$~T for the magnetic inverse-design task and $H\ge10$~GPa for the mechanical inverse-design task. 

The overall success rate was defined as the fraction of geometrically prescreened candidates that simultaneously satisfy compositional validity, uniqueness, novelty, thermodynamic stability, dynamical stability, and the prescribed target properties. This metric quantifies the effectiveness of the proposed inverse-design framework in identifying physically viable materials that meet the desired design objectives following geometric prescreening.

\section{Results and Discussion}

\subsection{Conditional fine-tuning from the pre-trained unconditional DiffCrysGen model}

All conditional models developed in this work are initialized from the pre-trained DiffCrysGen model. It was trained on a large-scale dataset comprising 411,940 crystalline materials curated from the Alexandria database~\cite{Alex-1, Alex-2, Alex-3}. The dataset consists of elementary, binary, and ternary compositions, with a maximum of 20 atoms per unit cell and an energy above the convex hull of $E_{\mathrm{hull}}\le0.5$~eV/atom. This relatively broad thermodynamic stability criterion was intentionally adopted during pre-training to maximize the diversity of the training data and to enable the diffusion model to learn a comprehensive representation of crystal chemical and structural space. By exposing the model to both thermodynamically stable and metastable crystal structures, the unconditional DiffCrysGen model learns a rich representation of the underlying structural distribution together with the fundamental crystallographic principles governing inorganic crystalline materials.

Starting from this unconditional model, we subsequently fine-tune the score network using labeled datasets. To demonstrate the generality of the proposed framework, we consider two representative multi-property inverse-design tasks: simultaneous conditioning on $h_{\mathrm{form}}$ and $M_\mathrm{s}$, and simultaneous conditioning on $h_{\mathrm{form}}$ and $H$. In both case studies, $h_{\mathrm{form}}$ is included as a proxy for thermodynamic stability, reflecting the fundamental requirement that candidate materials for any target application must be thermodynamically stable. These complementary tasks therefore demonstrate the applicability of the framework to both magnetic and mechanical materials design.

To systematically characterize the influence of CFG on the generative behaviour of the conditional models, we explored a broad range of guidance scales, $w\in\{0,1,\ldots,8\}$. At each guidance scale, 2,000 candidate crystal structures were generated for each inverse-design task by specifying the corresponding target property values. This systematic exploration enables us to quantify how the guidance strength influences target convergence, controllability, structural diversity, crystallographic symmetry, and the physical validity of the generated materials.

\subsection{Conditioning on formation energy and saturation magnetization}

\begin{figure*}[htbp]
    \centering
    
    {\raggedright \textsf{\textbf{(a)}} \par}
    \vspace{0.05cm} 
    \includegraphics[width=\linewidth]{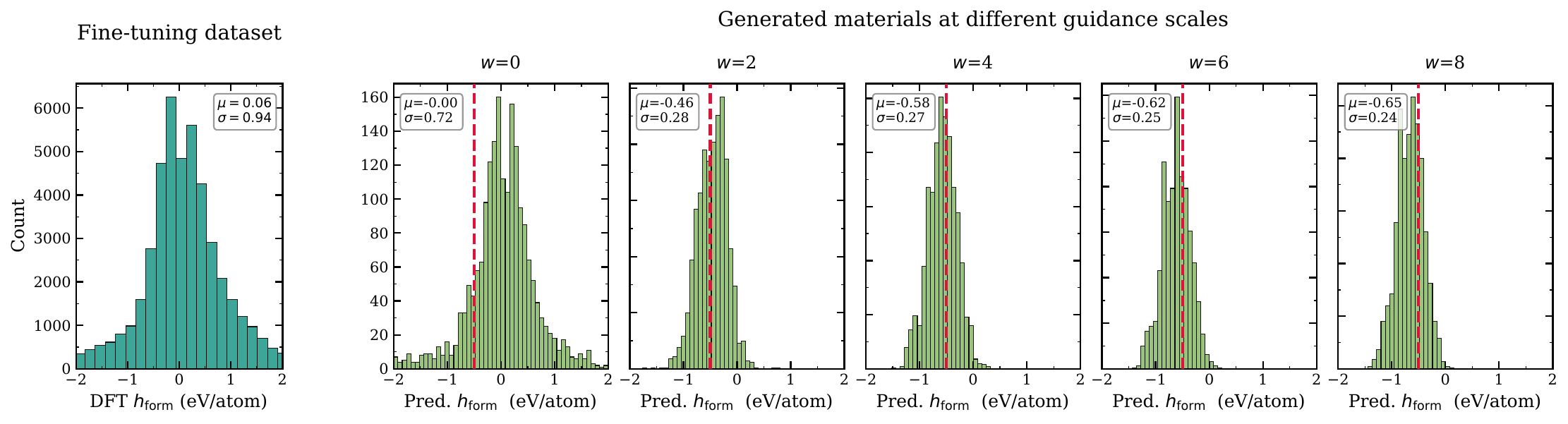}
    \label{fig:dist_hform}
    
    \vspace{0.1cm} 
    
    {\raggedright \textsf{\textbf{(b)}} \par}
    \vspace{0.05cm}
    \includegraphics[width=\linewidth]{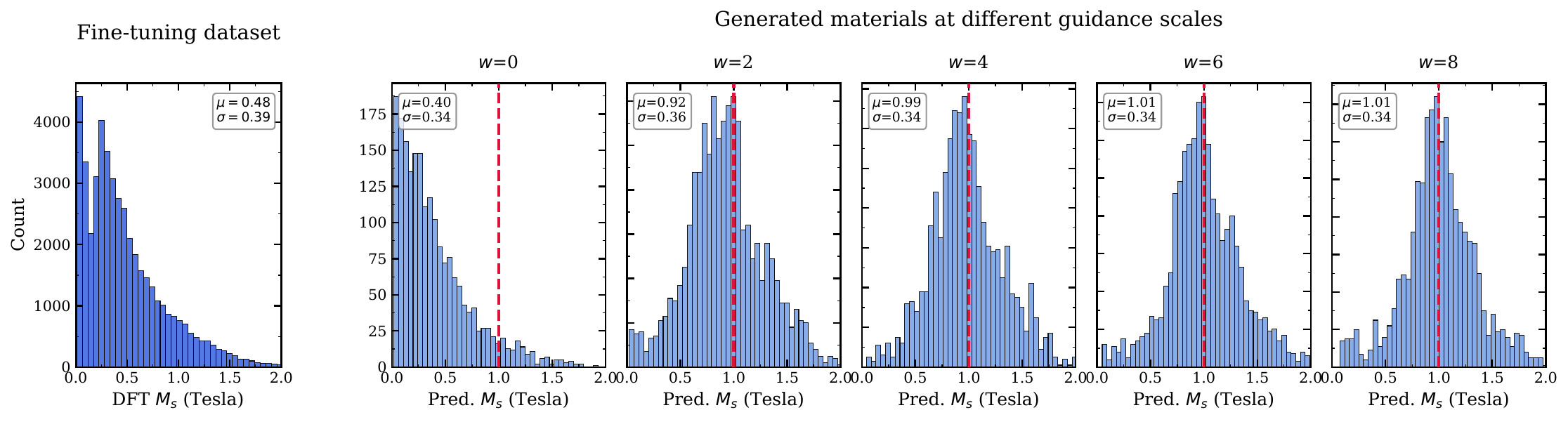}
    \label{fig:dist_ms}
    \caption{\textbf{Evolution of thermodynamic and functional property distributions under property guidance for $h_\mathrm{form}-M_s$ fine-tuned model.} (a) Distribution of DFT-calculated $h_{\mathrm{form}}$ for the fine-tuning dataset (left) and the predicted $h_{\mathrm{form}}$ distribution of the generated candidates (right) at different guidance scales $w$. (b) Corresponding distributions of DFT-calculated $M_s$ for the fine-tuning dataset (left) and the predicted $M_s$ distribution of the generated candidates (right). As the guidance scale increases, the generated distributions progressively shift toward the prescribed target values (dashed red lines).}
    \label{fig:combined_distributions}
\end{figure*}

The discovery of materials that simultaneously exhibit high $M_\mathrm{s}$ and thermodynamic stability is a longstanding objective in computational magnetic materials design. High-$M_\mathrm{s}$ materials, typically $M_s\ge1.0~$T, are essential for applications including permanent magnets, electric motors, and spintronic devices, while thermodynamic stability is a prerequisite for experimental synthesis and practical deployment. Consequently, the joint optimization of $h_\mathrm{form}$ and $M_\mathrm{s}$ represents a practically important inverse-design problem and serves as an ideal case study for evaluating the proposed property-guided framework.

As the first case study, we fine-tuned DiffCrysGen to simultaneously condition on $h_{\mathrm{form}}$ and $M_s$ using a curated labeled dataset of 46,496 materials from the Alexandria database. The fine-tuning dataset comprises materials upto $E_\mathrm{hull}\le0.5~$eV/atom with 25.5\% of the samples satisfying $E_{\mathrm{hull}}\le0.1$~eV/atom and 11.2\% exhibiting a high saturation magnetization ($M_s\ge1$~T). This intentionally imbalanced property distribution provides a stringent benchmark for evaluating the ability of the conditional model to generate materials satisfying target property combinations that are underrepresented in the fine-tuning    data. During sampling, the target properties were specified as $h_{\mathrm{form}}=-0.5$~eV/atom and $M_s=1.0$~T.

The effect of CFG on the generated property distributions is summarized in Fig.~\ref{fig:combined_distributions}. Increasing the guidance scale $w$ progressively steers both property distributions toward their prescribed targets while simultaneously reducing their variances, indicating increasingly precise control over the generated materials. For $h_\mathrm{form}$ (Fig.~\ref{fig:combined_distributions}a), the distribution mean shifts monotonically from $\mu=0$~eV/atom for the unconditional model ($w=0$) to $\mu=-0.65$~eV/atom at $w=8$, approaching the target value of $h_{\mathrm{form}}=-0.5$~eV/atom with a substantial reduction in the distribution width ($\sigma=0.72$ to $0.24$~eV/atom). Likewise, the $M_s$ distribution (Fig.~\ref{fig:combined_distributions}b) evolves from an unconditional mean of $\mu=0.40$~T to $\mu=1.01$~T at high guidance scales, exhibiting excellent convergence toward the prescribed target of $M_s=1.0$~T while maintaining a narrow distribution ($\sigma\approx0.34$~T). Notably, most of the target convergence is already achieved by intermediate guidance strengths ($w\approx4$), with only marginal changes observed at larger guidance scales, suggesting that the conditional model reaches a stable guidance regime without requiring excessively strong conditioning.

\begin{figure}[htbp]
\centering
\includegraphics[width=0.42\textwidth]{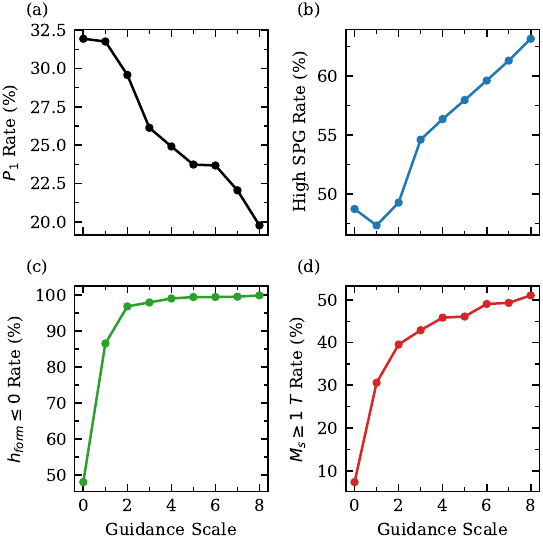}
\caption{\textbf{Effect of classifier-free guidance on multi-property crystal generation for $h_\mathrm{form}-M_s$ fine-tuned model}. Evolution of structural and functional metrics as a function of the guidance scale have been shown here. (a) The rate of generated structures with the lowest symmetry space group ($P_1$). (b) The percentage of generated structures with high-symmetry space groups. (c) The percentage of generated candidates with $h_{\mathrm{form}} \leq 0$. (d) Yield of targeted high-magnetization candidates with $M_s \geq 1\,\mathrm{T}$.}
\label{fig:cfg}
\end{figure}

A well-recognized limitation of crystal generative models is their tendency to overproduce structures belonging to the lowest-symmetry space group, $P1$,~\cite{Cheetham2024} in the absence of explicit symmetry constraints. Interestingly, we observe that CFG substantially mitigates this behaviour. As the guidance scale increases, the fraction of generated structures assigned to the $P1$ space group decreases monotonically from approximately 32\% for the unconditional model ($w=0$) to below 20\% at the highest guidance scale (Fig.~\ref{fig:cfg}a). Concurrently, the proportion of generated structures belonging to higher-symmetry space groups (space-group number $\ge16$, following the criterion introduced in DiffCrysGen~\cite{diffcrysgen}) increases from approximately 47\% to 63\% (Fig.~\ref{fig:cfg}b). This systematic shift toward higher crystallographic symmetry indicates that property guidance not only improves target-property controllability but also biases the generative process toward structurally more ordered crystal configurations. 

\begin{figure}[htbp]
\centering
\includegraphics[width=0.42\textwidth]{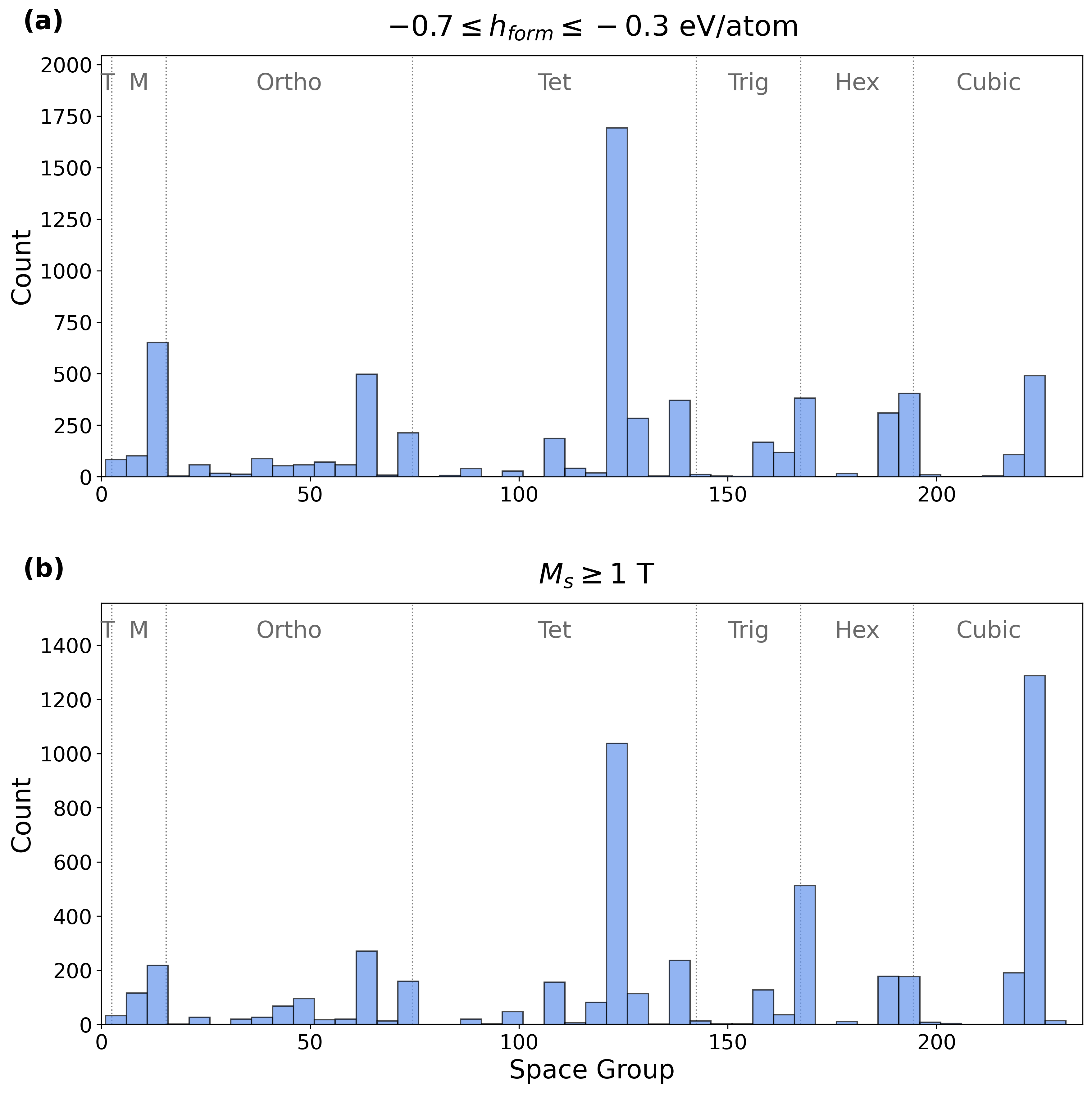}
\caption{\textbf{Space group distributions of materials in the fine-tuning dataset for the $h_{\mathrm{form}}-M_s$ fine-tuned model}. The data is partitioned to highlight the structural diversity within specific target regimes: (a)$-0.7 \leq h_{\mathrm{form}} \leq -0.3$~eV/atom, and (b) high saturation magnetization ($M_s \geq 1$~Tesla).}
\label{fig:train-ms}
\end{figure}

The observed structural evolution is presumably a consequence of the crystallographic characteristics of the target property regimes represented in the fine-tuning data. As shown in Fig.~\ref{fig:train-ms}, materials exhibiting high saturation magnetization ($M_s \ge 1$~T) span a broad range of crystallographic space groups with prominent peaks in tetragonal, hexagonal, and cubic crystal systems, indicating that high-$M_\mathrm{s}$ materials occupy structurally ordered regions of crystal chemical space. Consequently, the reduction of low-symmetry ($P1$) structures and the corresponding increase in higher-symmetry space groups observed under stronger guidance scale are consistent with the structural characteristics of the targeted magnetic materials present in the fine-tuning data.

In parallel, the functional yield improves consistently with increasing guidance strength. The fraction of generated materials satisfying $h_{\mathrm{form}}\le0$ increases from approximately 48\% at $w=0$ to nearly 100\% at $w=8$ (Fig.~\ref{fig:cfg}c), while the fraction of high-magnetization candidates ($M_s\ge1$~T) rises from  7.5\% to over 50\% (Fig.~\ref{fig:cfg}d). Notably, the largest improvements are achieved by intermediate guidance scales ($w\approx4$), beyond which there are only marginal gains. This behaviour suggests that an intermediate guidance window ($w=4$--$6$) provides an effective balance between achieving high functional yields and maintaining excellent crystallographic quality.

\begin{figure*}[htbp]
\centering
\includegraphics[width=0.75\textwidth]{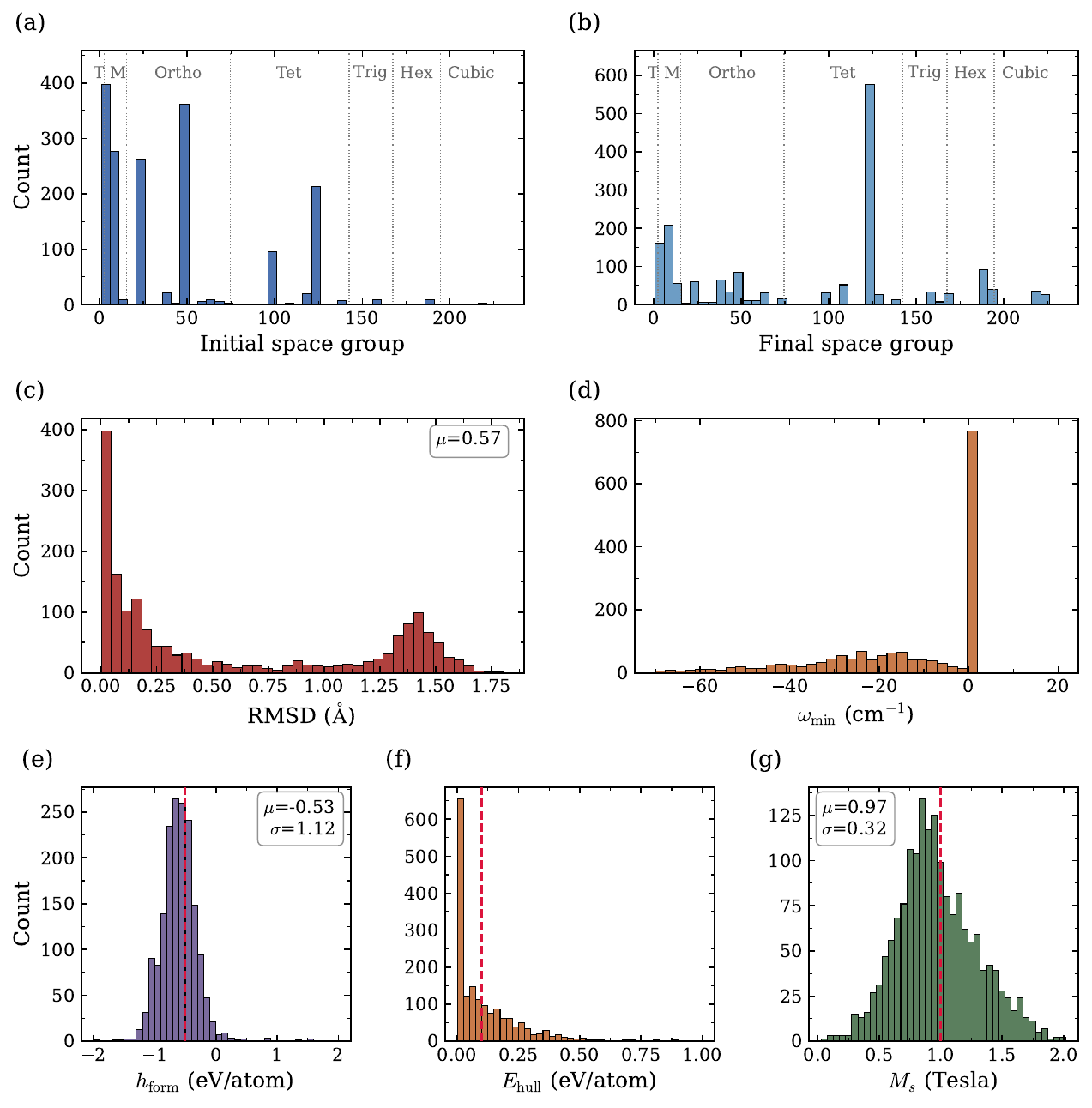}
\caption{\textbf{MLIP-based physical validation of the $h_\mathrm{form}-M_s$ fine-tuned model.} 
    (a) Initial space group distribution of the generated materials prior to relaxation. 
    (b) Final space group distribution after MLIP structural optimization. Vertical dotted lines in (a) and (b) demarcate the boundaries between major crystal systems. 
    (c) Distribution of RMSD between initial and MLIP-optimized structures, with a mean $\mu = 0.57$~\AA. 
    (d) Distribution of the minimum phonon frequency ($\omega_{\mathrm{min}}$), used to assess the dynamical stability of the optimized crystals. 
    (e) Distribution of $h_{\mathrm{form}}$. 
    (f) Distribution of $E_{\mathrm{hull}}$, evaluating thermodynamic phase stability. 
    (g) Distribution of the predicted $M_s$. 
    In panels (e), (f), and (g), the vertical dashed red lines indicate the targeted thresholds or ideal functional values (e.g., $M_s = 1.0$~T). Inset boxes display the distribution mean ($\mu$) and standard deviation ($\sigma$) where applicable.}
    \label{fig:mlip_opt_histograms}
\end{figure*}

To assess the physical viability of the 2,000 generated candidates, the structures generated at the guidance scale $w=4$ were first subjected to geometric prescreening to remove candidates containing unrealistically short interatomic distances. The remaining 1,709 candidates were subsequently processed using the proposed MLIP-based validation workflow. Figure~\ref{fig:mlip_opt_histograms} summarizes their structural, thermodynamic, dynamical, and magnetic characteristics following MLIP relaxation.

Fig.~\ref{fig:mlip_opt_histograms}a and Fig.~\ref{fig:mlip_opt_histograms}b compare the crystallographic space group distributions before and after MLIP relaxation. Structural optimization systematically shifts the generated crystals toward higher crystallographic symmetry. In particular, the fraction of structures assigned to the lowest-symmetry space group ($P1$) decreases markedly from 22.6\% to 8.2\%, while the population of tetragonal structures increases substantially after relaxation. At the same time, the relaxed structures continue to span a broad range of monoclinic, orthorhombic, tetragonal, trigonal, hexagonal, and cubic space groups, indicating that structural optimization enhances crystallographic symmetry. 

The structural refinement is further quantified by the distribution of root mean square displacement (RMSD) shown in Fig.~\ref{fig:mlip_opt_histograms}c. The distribution exhibits a bimodal character, with a large fraction of structures requiring only very small atomic displacements during relaxation, while a second population undergoes more substantial rearrangements before reaching local minima. Nevertheless, the mean RMSD remains only 0.57~\AA, indicating that the generated structures are, on average, already close to a local minima on the MLIP potential-energy surface. 

The dynamical stability of the relaxed structures was subsequently assessed using phonon band structure calculations. The distribution of the minimum phonon frequency ($\omega_\mathrm{min}$) is shown in Fig.~\ref{fig:mlip_opt_histograms}d. Based on the criterion $\omega_{\mathrm{min}}\ge-20~\mathrm{cm}^{-1}$, 65.5\% of the relaxed structures are classified as dynamically stable. 

The MLIP-predicted $h_\mathrm{form}$ distribution after relaxation is shown in Fig.~\ref{fig:mlip_opt_histograms}e, which exhibits excellent agreement with the prescribed target, with a mean value of $\mu=-0.53$~eV/atom, demonstrating that the conditional model successfully generates materials consistent with the requested thermodynamic design target.

The corresponding $E_{hull}$ distribution (Fig.~\ref{fig:mlip_opt_histograms}f) further demonstrates the thermodynamic stability of the generated candidates. Approximately 60.6\% of the relaxed structures satisfy $E_{\mathrm{hull}}\le0.1$~eV/atom, indicating a high proportion of potentially synthesizable materials.

Finally, the ML-predicted $M_s$ distribution after relaxation is shown in Fig.~\ref{fig:mlip_opt_histograms}g. The distribution remains in excellent agreement with the prescribed target, with a mean value of $0.97$~T, while $41.2\%$ of the relaxed structures satisfy the target criterion of $M_s\ge1$~T. These results demonstrate that the proposed inverse-design framework effectively preserves the desired magnetic functionality following structural relaxation.

A comprehensive summary of the MLIP-based validation results is presented in Table~\ref{tab:mlip_summary}. 
Among the geometrically prescreened candidates,  $44.5\%$ satisfy Stable--Unique--Novel (S.U.N.) criteria, while $12.3\%$ simultaneously satisfy compositional validity, uniqueness, novelty, thermodynamic stability, dynamical stability, and the prescribed magnetic target property. These results demonstrate that the proposed framework efficiently identifies physically viable magnetic materials while maintaining a high success rate under a stringent multi-criterion evaluation protocol.


\begin{figure*}[htbp]
\centering
\includegraphics[width=0.85\textwidth]{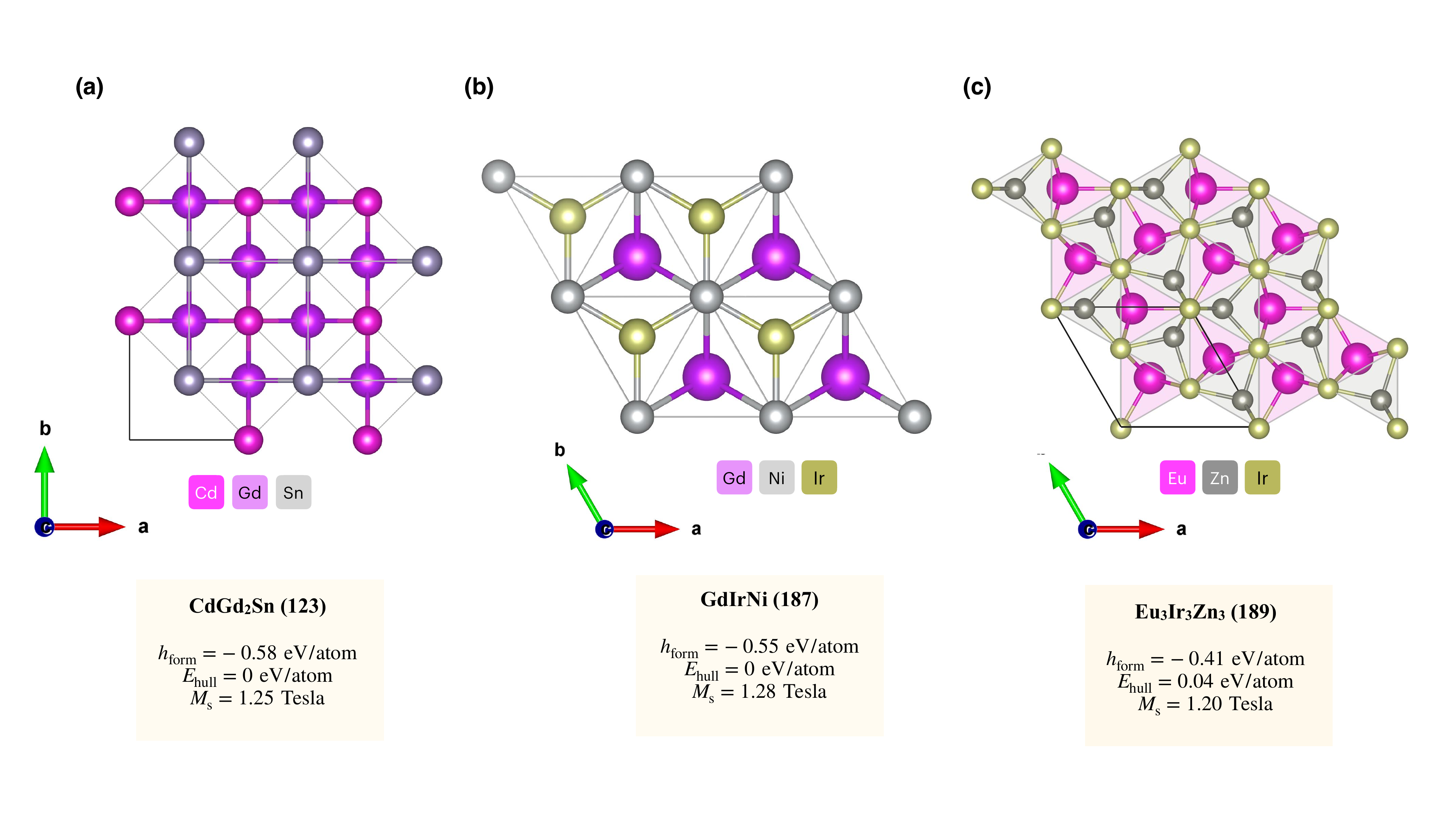}
\caption{\textbf{Representative magnetic materials discovered by the proposed inverse-design framework.} Shown here are three representative S.U.N. candidates: (a) CdGd$_2$Sn, (b) GdIrNi, and (c) Eu$_3$Ir$_3$Zn$_3$. All three materials satisfy the complete validation criteria, including structural validity, compositional validity, uniqueness, novelty, thermodynamic stability, dynamical stability, and the prescribed magnetic target ($M_s\ge1$~T). The numbers in parentheses denote the crystallographic space-group numbers.}    
\label{fig:ms_hform_materials}
\end{figure*}

\begin{table}[t]
\centering
\caption{\textbf{Summary of the MLIP-based validation results for the two inverse design tasks.} Reported percentages are evaluated for the geometrically prescreened candidates following MLIP structural relaxation. The S.U.N. rate denotes the fraction of candidates simultaneously satisfying thermodynamic stability, uniqueness, and novelty. The overall success rate denotes the fraction of geometrically prescreened candidates simultaneously satisfying compositional validity, uniqueness, novelty, thermodynamic stability ($E_{\mathrm{hull}}\le0.1$ eV/atom), dynamical stability ($\omega_{\mathrm{min}}\ge-20$ cm$^{-1}$), and the prescribed target property ($M_s\ge1$ T for the $h_{\mathrm{form}}$--$M_s$ model and $H\ge10$ GPa for the $h_{\mathrm{form}}$--$H$ model).}
\label{tab:mlip_summary}

\begin{tabular}{lcc}
\hline
\textbf{Metric} &
\textbf{$h_{\mathrm{form}}$--$M_s$} &
\textbf{$h_{\mathrm{form}}$--$H$} \\
\hline

\multicolumn{3}{l}{\textit{Candidate pool}}\\
Geometrically prescreened candidates & 1709 & 1318 \\
\hline

\multicolumn{3}{l}{\textit{Structural quality}}\\
Initial $P1$ rate (\%) & 22.6 & 18.6 \\
Final $P1$ rate (\%) & 8.2 & 7.9 \\
\hline

\multicolumn{3}{l}{\textit{Physical validation}}\\
Compositional validity (\%) & 93.4 & 80.6 \\
Uniqueness (\%) & 85.0 & 93.1 \\
Novelty (\%) & 94.0 & 68.5 \\
Thermodynamic stability (\%) & 60.6 & 41.3 \\
S.U.N. rate (\%) & 44.5 & 15.3 \\
Dynamical stability (\%) & 65.5 & 78.1 \\
Target-property satisfaction (\%) & 41.2 & 30.6 \\
\textbf{Overall success rate (\%)} & \textbf{12.3} & \textbf{3.9} \\
\hline

\end{tabular}
\end{table}


\subsection{Conditioning on formation energy and Vickers hardness}

As a second case study, we investigate the simultaneous optimization of $h_\mathrm{form}$ and $H$, a practically important inverse-design problem for the discovery of stable superhard materials. To estimate the macroscopic Vickers hardness ($H$), we employed the established empirical model~\cite{oganov2020}:
\begin{equation}
    H = \frac{0.096E (1 - 8.5\nu + 19.5\nu^2)}{1 - 7.5\nu + 12.2\nu^2 + 19.6\nu^3}
    \label{vickers_hardness}
\end{equation}
where $E$ denotes Young's modulus and $\nu$ represents Poisson's ratio. These intrinsic elastic properties are derived directly from the bulk modulus ($B$) and shear modulus ($G$) as follows:
\begin{equation}
    E=\frac{9BG}{3B+G}, ~~\nu = \frac{3B-2G}{6B+2G}
    \label{youngs_modulus}
\end{equation}

\begin{figure*}[htbp]
    \centering
    
    {\raggedright \textsf{\textbf{(a)}} \par}
    \vspace{0.05cm} 
    \includegraphics[width=\linewidth]{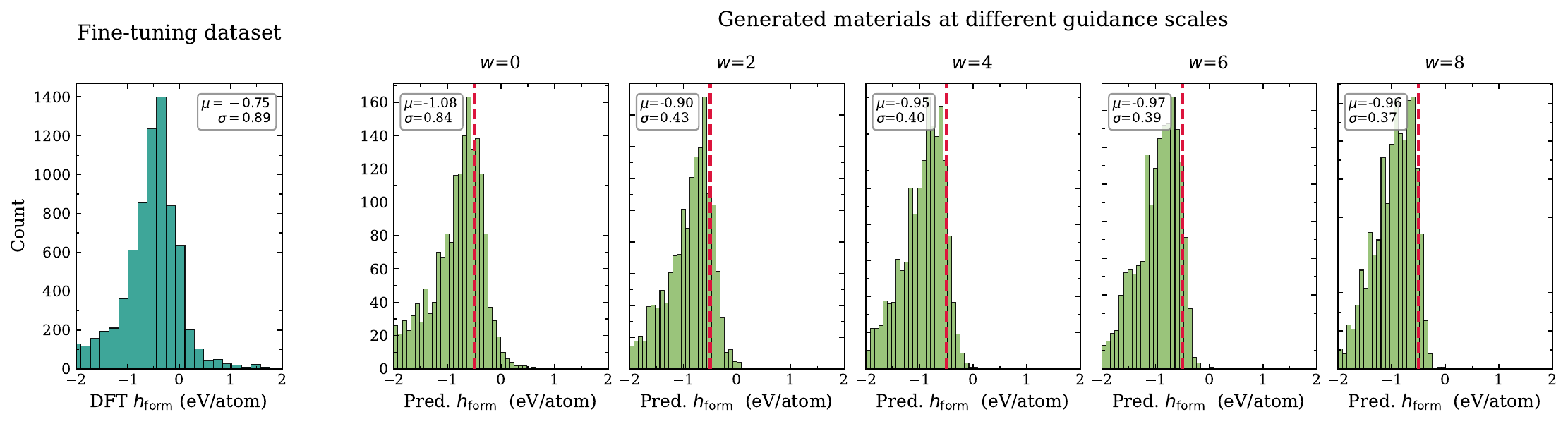}
    \label{fig:dist_hform}
    
    \vspace{0.1cm} 
    
    {\raggedright \textsf{\textbf{(b)}} \par}
    \vspace{0.05cm}
    \includegraphics[width=\linewidth]{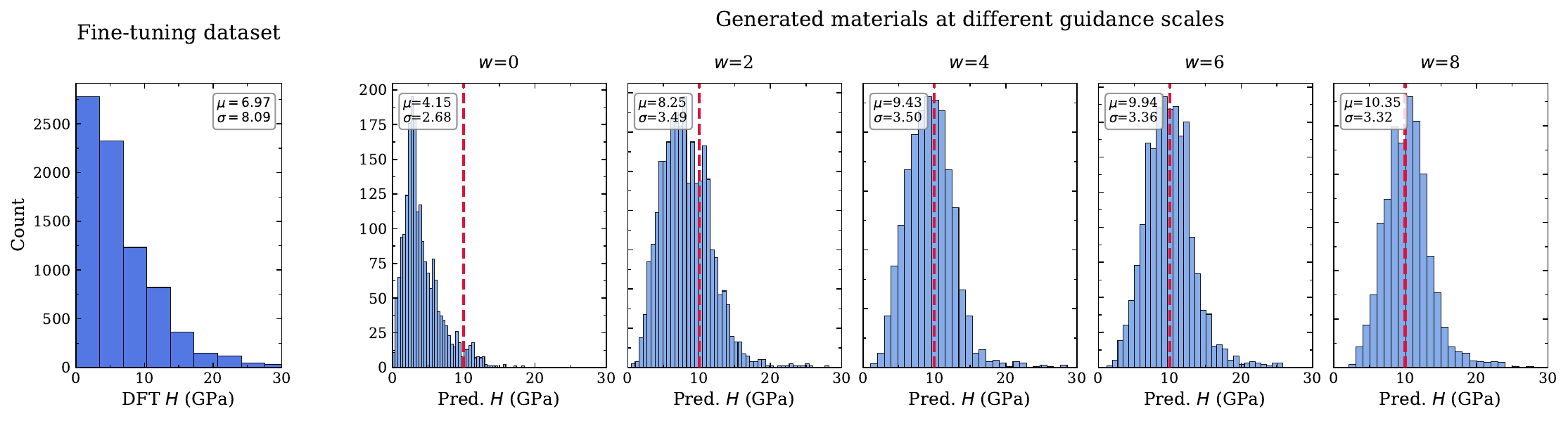}
    \label{fig:dist_vh}
    \caption{\textbf{Evolution of thermodynamic and functional property distributions under property guidance for $h_\mathrm{form}-H$ fine-tuned model.} (a) Distribution of DFT-calculated $h_{\mathrm{form}}$ for the fine-tuning dataset (left) and the predicted $h_{\mathrm{form}}$ distribution of the generated candidates (right) at different guidance scales $w$. (b) Corresponding distributions of DFT-calculated $H$ for the fine-tuning dataset (left) and the predicted $H$ distribution of the generated candidates (right). The dashed vertical lines represent the explicit design targets specified during generation.}
    \label{fig:combined_distributions_vh}
\end{figure*}

The dataset curated for this fine-tuning stage was sourced from the Materials Project~\cite{materials_project}, comprising 9,942 materials with explicitly reported values for both $B$ and $G$. Using these primary moduli, we calculated $E$ and $\nu$ (Eq.~\ref{youngs_modulus}), which were subsequently used to evaluate $H$ (Eq.~\ref{vickers_hardness}). The resulting training set distributions for $h_{\mathrm{form}}$ and $H$ are represented in Fig~\ref{fig:combined_distributions_vh}.

During conditional material generation, we specified simultaneous design targets of $h_{\mathrm{form}}=-0.5$~eV/atom and $H=10.0$~GPa. The hardness target of $10$~GPa was chosen as it represents the onset of the hard-material regime~\cite{oganov2020}, providing a practically relevant benchmark for evaluating the ability of the proposed framework to generate mechanically robust materials.

Figure~\ref{fig:combined_distributions_vh} summarizes the evolution of the generated property distributions with increasing guidance strength. Similar to the previous case study, CFG progressively narrows both distributions and improves adherence to the prescribed targets. However, unlike the $h_{\mathrm{form}}$--$M_s$ model, the $h_\mathrm{form}$ distribution does not converge exactly to the specified target (Fig.~\ref{fig:combined_distributions_vh}a). Instead, its mean shifts only from $-1.08$ to $-0.96$~eV/atom as the guidance scale increases. We attribute this behaviour to the intrinsic bias of the fine-tuning dataset, whose formation-energy distribution is itself strongly skewed toward lower energies ($\mu=-0.75$~eV/atom). Consequently, the unconditional structural prior favors highly stable materials, and CFG only partially shifts the generated distribution toward the specified target.

In contrast, the hardness distribution exhibits excellent controllability (Fig.~\ref{fig:combined_distributions_vh}b). Increasing the guidance scale continuously drives the generated distribution from an unconditional mean of 4.15~GPa toward the prescribed target, reaching a mean hardness of 10.35~GPa at $w=8$.

\begin{figure}[htbp]
\centering
\includegraphics[width=0.42\textwidth]{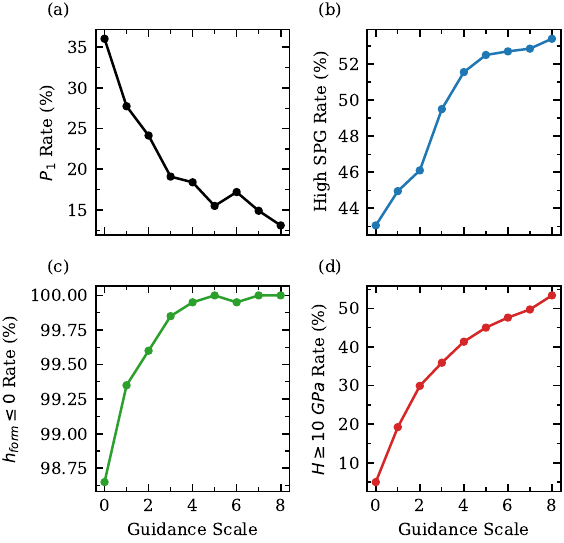}
\caption{\textbf{Effect of classifier-free guidance on multi-property crystal generation for $h_\mathrm{form}-H$ fine-tuned model}. Evolution of structural and functional metrics as a function of the guidance scale have been shown here. (a) The rate of generated structures with the lowest symmetry space group ($P_1$). (b) The percentage of generated structures with high-symmetry space groups. (c) The percentage of generated candidates with $h_{\mathrm{form}} \leq 0$. (d) Yield of targeted high Vickers hardness candidates with $H \geq 10\,\mathrm{GPa}$.}
\label{fig:cfg-vh}
\end{figure}

Importantly, this improved property control is again achieved without compromising crystallographic quality. The fraction of generated $P1$ structures decreases from approximately 36\% to 13\%, while the proportion of higher-symmetry space groups increases steadily with guidance (Fig.~\ref{fig:cfg-vh}a,b). 

\begin{figure}[htbp]
\centering
\includegraphics[width=0.42\textwidth]{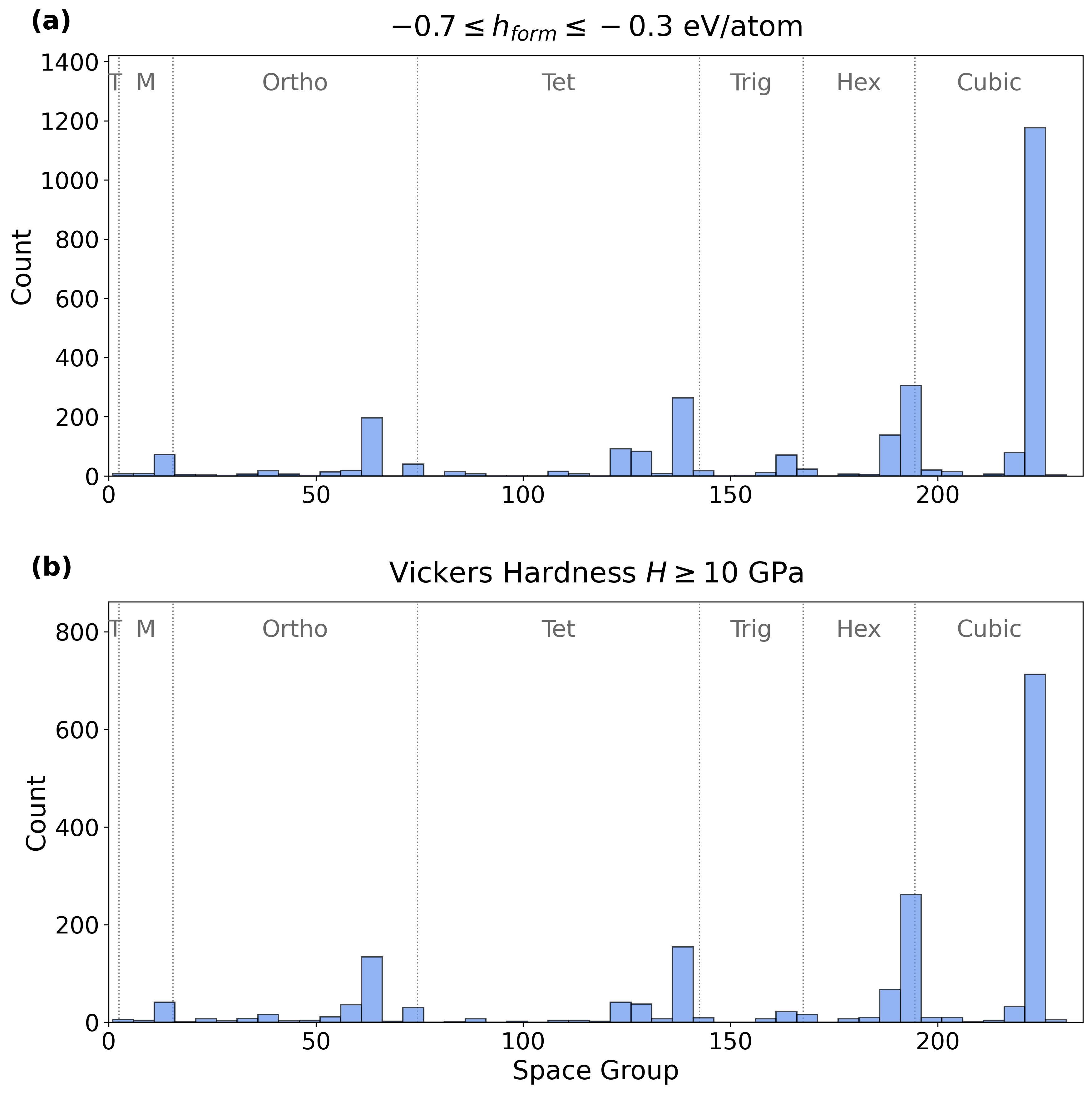}
\caption{\textbf{Space group distributions of the materials in fine-tuning dataset for the $h_{\mathrm{form}}-H$ fine-tuned model}. The data is partitioned to highlight the structural diversity within specific target regimes: (a)$-0.7 \leq h_{\mathrm{form}} \leq -0.3$~eV/atom, and (b) hard mechanical properties ($H \geq 10$~GPa).}
\label{fig:train-vh}
\end{figure}

This structural evolution demonstrates that the generative conditioning effectively captures a fundamental crystallographic principle: exceptional mechanical hardness and thermodynamic stability are intimately tied to dense, highly coordinated, and symmetric atomic packings. It is clearly reflected in the fine-tuning dataset (Fig.~\ref{fig:train-vh}), where the space group distributions for these target regimes are distinctly concentrated within high-symmetry classes, most notably the cubic, hexagonal, and tetragonal crystal systems.

Finally, the functional yield improves monotonically with increasing guidance strength. At intermediate guidance scales ($w=4$--$6$), more than 99.5\% of the generated structures satisfy $h_{\mathrm{form}}\le0$, while the fraction of superhard candidates ($H\ge10$~GPa) increases substantially (Fig.~\ref{fig:cfg-vh}c,d). These results demonstrate that the proposed framework generalizes beyond magnetic materials and enables efficient inverse design across distinct classes of target properties.

\begin{figure*}[htbp]
\centering
\includegraphics[width=0.75\textwidth]{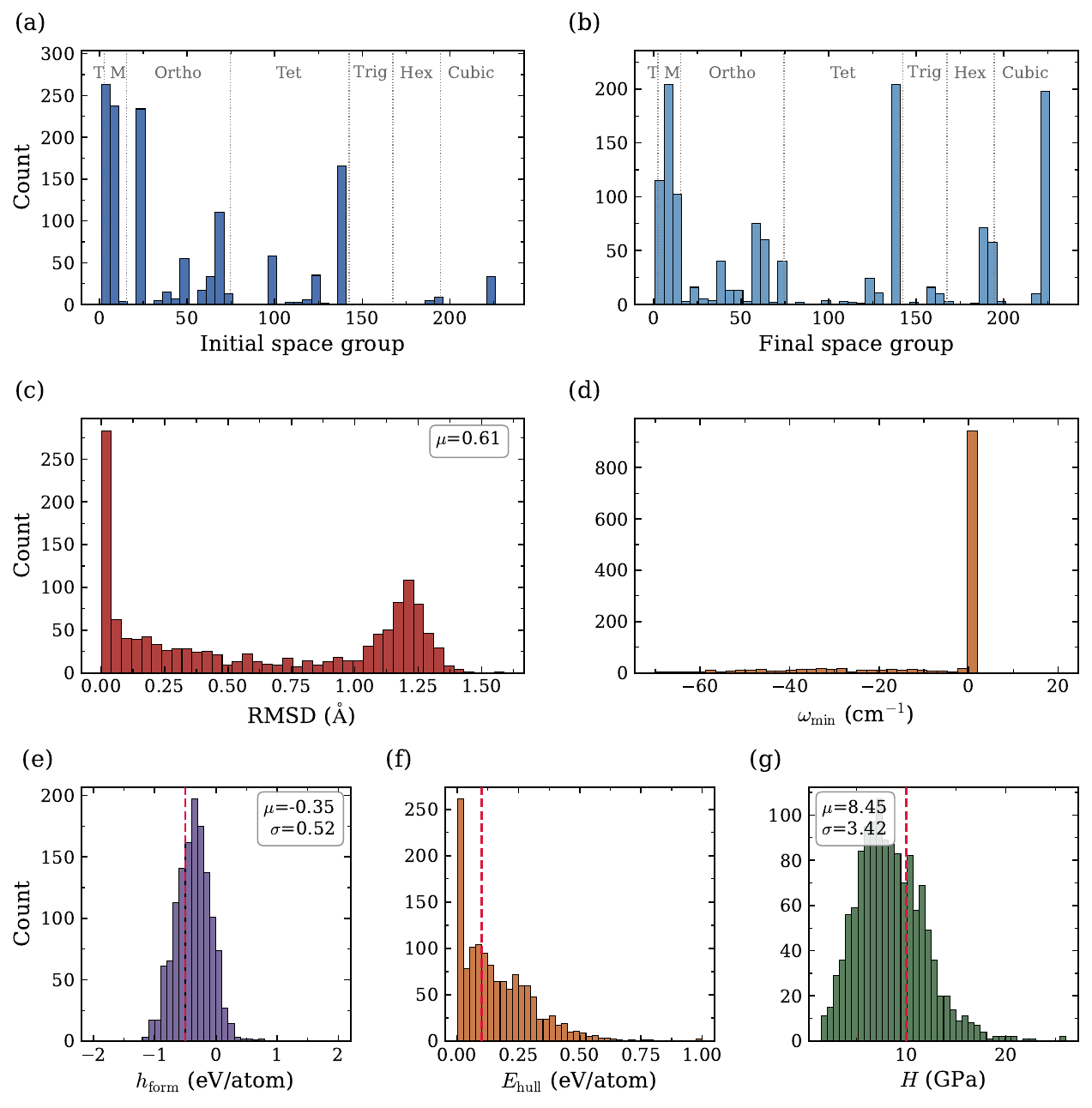}
\caption{\textbf{MLIP-based physical validation of the $h_\mathrm{form}-H$ fine-tuned model.} 
    (a) Initial space group distribution of the generated materials prior to relaxation. 
    (b) Final space group distribution after MLIP structural optimization. Vertical dotted lines in (a) and (b) demarcate the boundaries between major crystal systems. 
    (c) Distribution of RMSD between initial and MLIP-optimized structures, with a mean $\mu = 0.57$~\AA. 
    (d) Distribution of the minimum phonon frequency ($\omega_{\mathrm{min}}$), used to assess the dynamical stability of the optimized crystals. 
    (e) Distribution of $h_{\mathrm{form}}$. 
    (f) Distribution of $E_{\mathrm{hull}}$,
    (g) Distribution of the predicted $H$. 
    In panels (e), (f), and (g), the vertical dashed red lines indicate the targeted thresholds or ideal functional values (e.g., $H = 10$~GPa). Inset boxes display the distribution mean ($\mu$) and standard deviation ($\sigma$) where applicable.}    \label{fig:mlip_opt_histograms_vh}
\end{figure*}

To assess the physical viability of the generated hard-material candidates, the 1,318 structures retained after geometric pre-screening from the generation at guidance scale $w=4$ were subjected to the proposed MLIP-based validation workflow. The resulting structural, dynamical, thermodynamic, and mechanical characteristics are summarized in Fig.~\ref{fig:mlip_opt_histograms_vh}. 

Figures~\ref{fig:mlip_opt_histograms_vh}a and \ref{fig:mlip_opt_histograms_vh}b compare the crystallographic space group distributions before and after MLIP relaxation. Similar to the $h_{\mathrm{form}}$--$M_s$ case study, structural optimization systematically improves the crystallographic quality of the generated structures, reducing the fraction of structures belonging to the lowest-symmetry space group ($P1$) from 18.6\% to 7.9\%. At the same time, the optimized structures continue to span a broad range of crystal symmetries, indicating that structural relaxation enhances crystallographic symmetry while preserving structural diversity.

The structural refinement is further quantified by the RMSD distribution shown in Fig.~\ref{fig:mlip_opt_histograms_vh}c. The mean RMSD is only 0.61~\AA, indicating that the generated structures are, on average, already close to local minima on the MLIP potential-energy surface.

The optimized structures also exhibit excellent stability. Using the practical criterion of $\omega_{\mathrm{min}}\ge-20~\mathrm{cm}^{-1}$, 78.1\% of the candidates are dynamically stable (Fig.~\ref{fig:mlip_opt_histograms_vh}d). Furthermore, 90.2\% possess negative formation energies (Fig.~\ref{fig:mlip_opt_histograms_vh}e), while 41.3\% satisfy $E_{\mathrm{hull}}\le0.1$~eV/atom (Fig.~\ref{fig:mlip_opt_histograms_vh}f). The optimized structures also retain excellent mechanical performance, with a mean Vickers hardness of 8.45~GPa and 30.6\% satisfying the target criterion of $H\ge10$~GPa (Fig.~\ref{fig:mlip_opt_histograms_vh}g).

A comprehensive summary of the MLIP-based validation results is presented in Table~\ref{tab:mlip_summary}. Among the geometrically prescreened candidates, 15.3\% satisfy the S.U.N. criteria, while 3.9\% simultaneously satisfy compositional validity, uniqueness, novelty, thermodynamic stability, dynamical stability, and the prescribed hardness target property. These results demonstrate that the proposed framework generalizes beyond magnetic materials and effectively identifies physically viable hard materials under the same stringent multi-stage evaluation protocol, providing a promising pool of candidates for subsequent first-principles validation.

\begin{figure*}[htbp]
\centering
\includegraphics[width=0.85\textwidth]{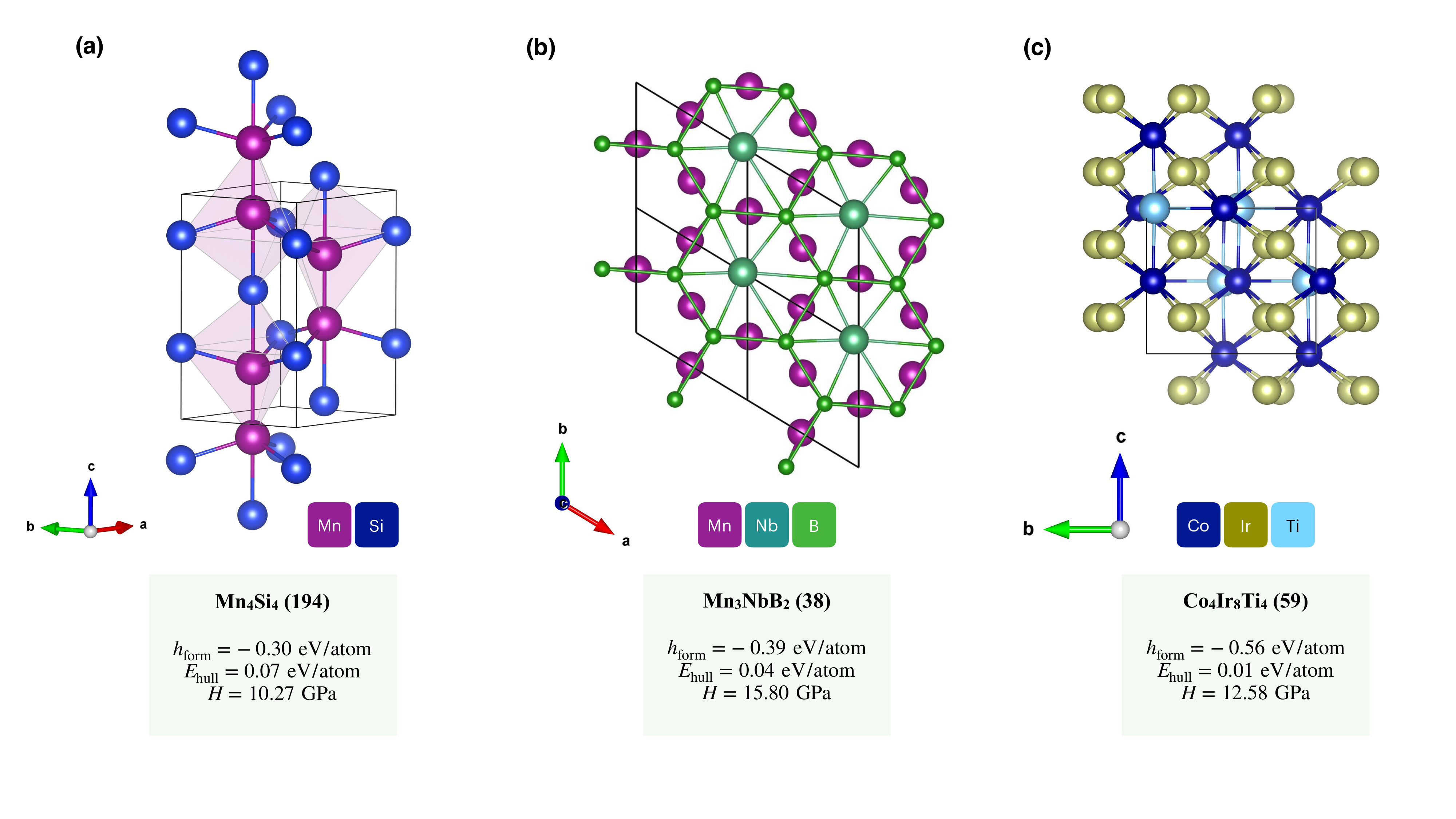}
\caption{\textbf{Representative hard materials discovered by the proposed inverse-design framework.} Shown here are three S.U.N. candidates: (a) Mn$_4$Si$_4$, (b) Mn$_3$NbB$_2$, and (c) Co$_4$Ir$_8$Ti$_4$. All three materials satisfy the complete validation criteria, including structural validity, compositional validity, uniqueness, novelty, thermodynamic stability, dynamical stability, and the prescribed hardness target ($H\ge10$~GPa). The numbers in parentheses denote the crystallographic space-group numbers.
}
    \label{fig:ms_hform_materials}
\end{figure*}


\section{Conclusion}

We have established an efficient framework for inverse design of crystalline materials by integrating adapter-based fine-tuning and classifier-free guidance into the lightweight DiffCrysGen diffusion model. The proposed framework efficiently adapts a pre-trained unconditional diffusion model to diverse multi-property generation tasks while preserving its computational efficiency. With a raw sampling speed of approximately 115 crystal structures per second, the framework enables rapid large-scale exploration of the crystal design space, where computationally inexpensive geometric prescreening efficiently removes physically unrealistic candidates before computationally intensive MLIP-based validation.

Beyond developing a property-guided generative model, this work provides a systematic understanding of how classifier-free guidance influences crystal generation. By investigating a broad range of guidance scales, we show that increasing the guidance strength progressively improves target convergence and functional yield while simultaneously enhancing crystallographic quality through a reduction of low-symmetry ($P1$) structures and an increase in higher-symmetry crystal structures. These observations demonstrate that controllable crystal generation need not compromise structural realism, thereby addressing a fundamental question surrounding property-guided diffusion models.

The generality of the proposed framework is demonstrated through two distinct inverse-design tasks targeting magnetic and mechanical materials. Across both case studies, the generated candidates exhibit excellent agreement with the prescribed property targets while maintaining broad crystallographic diversity. Subsequent MLIP-based structural optimization, phonon calculations, thermodynamic stability analysis, and property evaluation further confirm that a substantial fraction of the generated materials remain physically viable.

The proposed framework is inherently modular and readily extensible to arbitrary combinations of continuous material properties without modification of the underlying diffusion architecture. More broadly, this work establishes that lightweight, high-throughput diffusion models can serve as practical and scalable engines for controllable inverse materials design, providing a foundation for accelerating the discovery of functional crystalline materials across diverse application domains.

\section{Acknowledgements}

We acknowledge the High-Performance Computing (HPC) facility at the Harish-Chandra Research Institute (HRI), Prayagraj, India, for the computational resources used in this work (\url{https://www.hri.res.in/cluster/}). All machine learning model training and related computations were carried out using the NVIDIA H100 GPU available through the HRI HPC cluster.

\bibliography{ref}

\end{document}